\documentclass[letterpaper]{article} 
\usepackage{aaai2026}  
\usepackage{times}  
\usepackage{helvet}  
\usepackage{courier}  
\usepackage[hyphens]{url}  
\usepackage{graphicx} 
\usepackage{natbib}  
\usepackage{caption} 
\usepackage{algorithm}
\usepackage{algorithmic}
\usepackage{amssymb}
\usepackage{amsmath}
\usepackage{booktabs}
\usepackage{multirow} 
\usepackage{newfloat}
\usepackage{listings}
\DeclareCaptionStyle{ruled}{labelfont=normalfont,labelsep=colon,strut=off} 
\floatstyle{ruled}
\newfloat{listing}{tb}{lst}{}
\floatname{listing}{Listing}
\title{Forgetting by Pruning: Data Deletion in Join Cardinality Estimation}

\author {
    Chaowei He\textsuperscript{\rm 1},
    Yuanjun Liu\textsuperscript{\rm 1},
    Qingzhi Ma\textsuperscript{\rm 1}\thanks{Corresponding author},
	Shenyuan Ren\textsuperscript{\rm 2},
	Xizhao Luo\textsuperscript{\rm 1},
	Lei Zhao\textsuperscript{\rm 1},
	An Liu\textsuperscript{\rm 1}\footnotemark[1]
}
\affiliations {
    \textsuperscript{\rm 1}Soochow University\\
    \textsuperscript{\rm 2}Beijing Jiaotong University\\
    cwhe@stu.suda.edu.cn, yjliu1@stu.suda.edu.cn, qzma@suda.edu.cn, syren@bjtu.edu.cn, xzluo@suda.edu.cn, zhaol@suda.edu.cn, anliu@suda.edu.cn
}

\begin{document}

\maketitle

\begin{abstract}
	Machine unlearning in learned cardinality estimation (CE) systems presents unique challenges due to the complex distributional dependencies in multi-table relational data. Specifically, data deletion, a core component of machine unlearning, faces three critical challenges in learned CE models: attribute-level sensitivity, inter-table propagation and domain disappearance leading to severe overestimation in multi-way joins. We propose Cardinality Estimation Pruning (CEP), the first unlearning framework specifically designed for multi-table learned CE systems. CEP introduces Distribution Sensitivity Pruning, which constructs semi-join deletion results and computes sensitivity scores to guide parameter pruning, and Domain Pruning, which removes support for value domains entirely eliminated by deletion. We evaluate CEP on state-of-the-art architectures NeuroCard and FACE across IMDB and TPC-H datasets. Results demonstrate CEP consistently achieves the lowest Q-error in multi-table scenarios, particularly under high deletion ratios, often outperforming full retraining. Furthermore, CEP significantly reduces convergence iterations, incurring negligible computational overhead of 0.3\%-2.5\% of fine-tuning time.
\end{abstract}
\begin{links}
	\link{Code}{https://github.com/heriec/CEP}
\end{links}

\section{Introduction}

Machine unlearning \cite{cao2015towards, guo2019certified, bourtoule2021machine}, the task of removing the influence of specific data points from trained models, has become increasingly important for data privacy and dynamic model updates. Regulations such as GDPR \cite{voigt2017gdpr} and CCPA \cite{goldman2020introduction} mandate the right to data deletion, while real-world systems often require models to adapt to deletions from data expiration or correction. In database (DB) systems, machine learning (ML) models are increasingly being explored to enhance core components. Among them, cardinality estimation (CE), a fundamental task in query optimization, must reflect data deletions accurately to maintain query performance. This necessitates applying unlearning techniques to ensure the model's predictions remain reliable after data removal.

Unfortunately, current learned CE systems lack effective mechanisms for handling data deletions. Existing approaches either require expensive full model retraining for every deletion or resort to naive fine-tuning that ignores the distributional shifts in relational data. Recent work like \citet{kurmanji2024machine} applies unlearning to downstream DB tasks but focuses only on single-table scenarios, neglecting the complexities of multi-table join estimation. Moreover, this work relies on existing unlearning methods without introducing new strategies tailored to the specific challenges of cardinality estimation.

The fundamental challenge lies in CE's unique characteristics that distinguish it from traditional machine learning scenarios. First, \textbf{attribute-level sensitivity} varies dramatically based on value rarity and distribution patterns. For instance, deleting 1,000 ``Action'' movies from a large dataset causes minimal distributional impact since many similar records remain, while deleting just 5 ``Film-Noir'' movies may eliminate the entire genre, drastically altering the model's learned representations. Second, \textbf{inter-table propagation} compounds this complexity, as deletions cascade through foreign key relationships, affecting not only the target table but also related tables and their join distributions. A single actor deletion impacts actor-specific queries, movie-actor join cardinalities, and potentially multi-way joins involving genres, directors, and ratings.

Another critical challenge is \textbf{domain disappearance}, where deletions remove all instances of certain attribute values, shrinking the model's input space. This phenomenon is particularly problematic in multi-table scenarios where join operations amplify estimation errors. Without proper handling, models continue to assign non-zero probabilities to vanished domains, leading to severe overestimation in complex multi-way joins. Existing unlearning methods, designed for independent data points, cannot address such domain-level changes inherent in structured relational data.

To address these challenges that remain unaddressed by current unlearning methods in learned CE, we propose \textbf{Cardinality Estimation Pruning (CEP)}, an unlearning framework specifically designed for multi-table CE tasks. CEP introduces two key components:
(1) \textbf{Distribution Sensitivity Pruning} addresses the core challenge of attribute-level sensitivity and inter-table propagation. It constructs semi-join deletion results using join keys and computes sensitivity scores from distributional shifts to guide fine-grained parameter pruning.
(2) \textbf{Domain Pruning} tackles domain disappearance by removing deleted value domains from the model's input space, ensuring zero probability mass is assigned to obsolete values.
Following the pruning phase, CEP performs fine-tuning on the retained data to restore model performance and ensure accurate CE on the retained dataset.

In summary, our contributions are as follows:
\begin{itemize}
	\item We present the first unlearning framework for multi-table learned CE, identifying key challenges in relational data deletion and proposing CEP with specialized pruning techniques to enable efficient removal of training records without full retraining.
	\item We design two specialized components:
	      (i) Distribution Sensitivity Pruning computes sensitivity scores from distributional shifts between full and retained join results to guide fine-grained parameter pruning;
	      (ii) Domain Pruning removes support for value domains entirely eliminated by deletion, ensuring no residual probability mass.
	\item We evaluate CEP on NeuroCard and FACE across IMDB and TPC-H datasets, demonstrating CEP consistently achieves the lowest Q-error in multi-table scenarios, particularly under high deletion ratios, often outperforming even full retraining. Furthermore, CEP significantly reduces convergence iterations, incurring negligible computational overhead (0.3\%-2.5\% of fine-tuning time).
\end{itemize}

\begin{figure*}[t]
	\centering
	\includegraphics[width=0.97\textwidth]{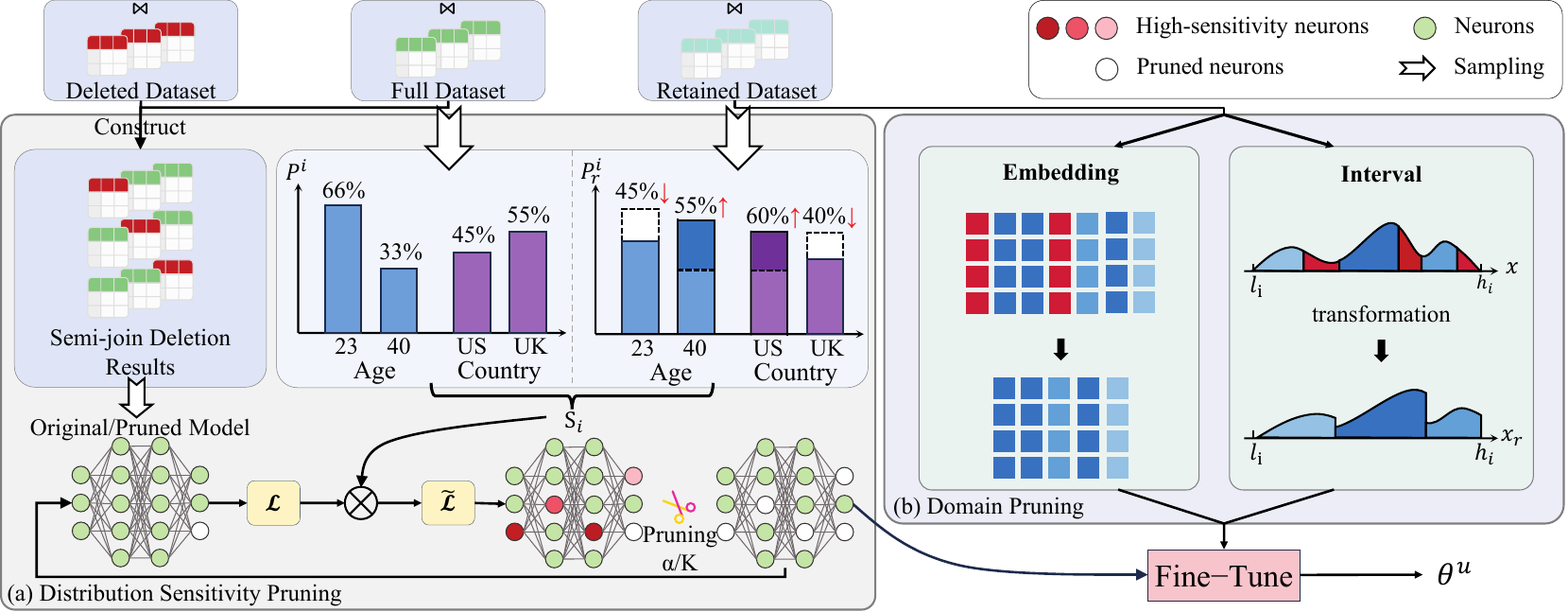}
	\caption{
		Overview of our proposed method \textbf{Cardinality Estimation Pruning (CEP)},
		(a) \textbf{Distribution Sensitivity Pruning}: constructs semi-join deletion results and computes sensitivity scores from distributional shifts between the full and retained join results. These weights are used to identify and prune highly sensitive parameters.
		(b) \textbf{Domain Pruning}: removes support for value domains entirely erased by the deletion.
		Pruned models are then fine-tuned on the retained dataset to restore performance.
	}

	\label{overview}
\end{figure*}

\section{Related Work}

\subsection{Machine Learning for CE}

Recent research integrates ML into database systems, with CE being a prominent focus. Traditional methods based on histograms \cite{matias1998wavelet, poosala1996improved} and sampling \cite{olken1993random, haas1995sampling} often perform poorly on complex queries such as multi-table joins due to attribute independence assumptions. Neural networks, known for their expressive power, have been proposed to overcome this. Naru \cite{yang2019deep} uses autoregressive modeling for conditional distributions; NeuroCard \cite{yang2020neurocard} extends this to multi-table joins by modeling joint distributions; and FACE \cite{wang2021face} adopts normalizing flows to learn invertible transformations, improving estimation accuracy and generalization. However, these models inherently assume static data distributions, making them ill-equipped to handle dynamic data changes, particularly data deletions, vital for compliance and real-world database management.

To handle dynamic data, recent efforts have explored adapting CE models to evolving data. Query-driven methods like Warper \cite{li2022warper} and CardOOD \cite{li2024cardood} adapt to evolving workloads through continual query feedback. While valuable, these approaches rely on specific query patterns or extensive historical logs. Data-driven approaches for dynamic CE have also begun to emerge: \citet{kurmanji2023detect} investigate learning under data insertions via transfer learning. More specifically concerning data deletion and unlearning, \citet{kurmanji2024machine} directly addresses data deletions in CE. Crucially, however, this work focuses exclusively on single-table scenarios, leaving the complex challenges of multi-table join estimation for unlearning entirely unaddressed. This multi-table limitation is precisely what our work tackles.

\subsection{Machine Unlearning}

Machine unlearning in deep models aims to remove the influence of specific data points from a trained model without full retraining. Early approaches include \citet{wu2020deltagrad}, which leveraged stored gradient information for efficient approximation of model updates after data deletion. Influence function-based methods, pioneered by \citet{guo2019certified}, estimate individual data point impact, but often incur high computational costs due to expensive second-order Hessian computations \cite{basu2020influence} and perform poorly on non-convex models.
To address computational overhead, several works approximate the Hessian. \citet{golatkar2020eternal} approximated it using the Fisher information matrix (FIM), enabling more tractable updates. Building on this, \citet{liu2023unlearning} further leveraged FIM to derive theoretical bounds on model divergence via influence functions. Another notable FIM-based approach, Selective Synaptic Dampening (SSD) by \citet{foster2024fast}, identifies and suppresses parameters most influenced by forgotten samples by comparing FIMs from retain and deletion data.

Beyond gradient and influence-based techniques, other unlearning strategies involve structural changes or knowledge distillation. \citet{jia2023model} showed that model sparsity can improve unlearning efficacy. SCRUB \cite{kurmanji2023towards} adopts a teacher-student framework to minimize alignment with deleted data while maximizing retention of retained data. While these diverse methods advance machine unlearning, they largely focus on classification and give limited attention to regression settings. \citet{tarun2023deep} proposed Blindspot Unlearning by fine-tuning with Gaussian noise replaced deletions, and \citet{chen2025Regression} extended influence function from linear to logistic regression. Crucially, none of these existing unlearning techniques address the intricate attribute-level sensitivity, inter-table propagation, or domain disappearance unique to multi-table relational data in cardinality estimation. Their general-purpose nature lacks specialized mechanisms for accurate and efficient unlearning in CE models, motivating our novel CEP framework.

\section{Preliminaries}

We consider a relational database $\mathcal{D}$ consisting of multiple tables ${T^1, T^2, \dots, T^n}$, where each table $T$ contains tuples defined over a set of attributes ${A^1, A^2, \dots, A^m}$. Each attribute $A^i$ has an associated domain $Dom(A^i)$, representing the set of distinct values it can take.

Given a pre-trained model with parameters $\theta^o$ on the full(i.e. original) dataset $\mathcal{D}$, we aim to remove the influence of a deleted dataset $\mathcal{D}_d \subset \mathcal{D}$ while preserving performance on the retain set $\mathcal{D}_r = \mathcal{D} \setminus \mathcal{D}_d$. In multi-table scenarios, deletions occur per table: for each table $T^i$, we have retained and deleted subsets $T_r^i$ and $T_d^i$ respectively, where $\mathcal{D}_r = {T_r^1, \ldots, T_r^n}$ and $\mathcal{D}_d = {T_d^1, \ldots, T_d^n}$.
We denote the model's cardinality estimate of a query $q$ under original parameters $\theta^o$ as $|C| = f(q; \theta^o) \cdot |T|$, where $f(\cdot; \theta^o)$ is a selectivity estimator learned by minimizing a loss function $\mathcal{L}$ and $|T|$ is the cardinality of the full outer join. The goal of unlearning is to update parameters $\theta^o$ to $\theta^u$, such that it behaves as if $\mathcal{D}_d$ never exists.

\section{Cardinality Estimation Pruning}

In this section, we present CEP, which integrates distribution sensitivity pruning and domain pruning to effectively remove the influence of deleted data from CE models. The overall pipeline is shown in Figure \ref{overview}.

\subsection{Distribution Sensitivity Pruning}

To effectively prune models in response to data deletions, we propose Distribution Sensitivity Pruning, which measures each model parameter's sensitivity to deleted data and prunes highly sensitive ones.

\subsubsection{Sensitivity Scores.}

Our goal is to compute sensitivity scores that accurately reflect the impact of deleted data on model parameters. While methods like the Hessian matrix or its first-order approximation, the Fisher Information Matrix (FIM), are commonly used to assess parameter importance, their computational cost can be prohibitive in practice. To ensure scalability, we adopt the diagonal approximation of the FIM. This significantly reduces both computational and memory overhead by approximating the importance of parameter $\theta$ with its squared gradient magnitude over the deleted dataset $\mathcal{D}_d$:
\begin{equation}
	\mathcal{I}(\mathcal{L},\mathcal{D}_d)=\mathbb{E}_{x\in \mathcal{D}_d}\left[\left(\frac{\partial\mathcal{L}(x)}{\partial\theta}\right)^2\right]
	\label{eq:fim}
\end{equation}
While $\mathcal{I}(\mathcal{L},\mathcal{D}_d)$ provides a general measure of parameter importance, they don't inherently capture the attribute-level sensitivity crucial for data deletion. To address this, we introduce a novel attribute sensitivity measure $S_{i}$, designed to quantify how deletions impact the distributions of individual attributes. From an information-theoretic perspective, $S_{i}$ quantifies the distributional shift between the original and retained attribute values, evealing significant information changes. For each attribute $A^i$, we define $S_{i}$ as:
\begin{equation}
	S_{i}=\frac{|P^i-P_r^i|}{P_r^i}
\end{equation}
where $P^i$ and $P_r^i$ denote the empirical probability mass functions (or histograms for categorical attributes) of attribute $A^i$ in the full and retained datasets, respectively. $S_i$ measures the relative distributional shift of each attribute, guiding the model to focus pruning on parameters most affected by deletion. We incorporate $S_i$ into the loss function to prioritize attributes with larger shifts, yielding a modified loss $\tilde{\mathcal{L}}(x)$ tailored to the model architecture.
\begin{itemize}
	\item \textbf{Autoregressive Models:} Since AR models learn conditional probability  distributions $p(A^i|A^{<i})$, we apply $S_{i}$ directly to each conditional term: $\tilde{\mathcal{L}}(x)=-\sum_{i=1}^DS_i\cdot\log p(A^i\mid A^{<i})$.
	\item \textbf{Normalizing Flow Models:} For NF models that learn joint probability distributions through invertible transformations, we aggregate column shifts across all attributes present in a sample: $\tilde{\mathcal{L}}(x)=\mathcal{L}(x)\cdot\left(\sum_{i\in\mathrm{Cols}(x)}S_i\right)$
\end{itemize}
The final sensitivity scores is computed as:
\begin{equation}
	\mathcal{I}(\tilde{\mathcal{L}},\mathcal{D}_d)=\mathbb{E}_{x\in \mathcal{D}_d}\left[\left(\frac{\partial\tilde{\mathcal{L}}(x)}{\partial\theta}\right)^2\right]
\end{equation}

\subsubsection{Multi-Table Pruning Strategy.}

In multi-table scenarios, individual table statistics may not accurately reflect the global distributional changes after join operations. To address this challenge, we need to estimate how data deletion from each table affects the final joined result. We construct \textbf{semi-join deletion results}  $\{J_d^k\}_{k=1}^K$ to simulate the impact of deleted data on the overall join distribution. For each table $T^k$, we compute:
\begin{equation}
	J_d^k = T^1 \bowtie \cdots \bowtie T^{k-1} \bowtie T_d^k \bowtie T^{k+1} \bowtie \cdots \bowtie T^n
\end{equation}
where $T_d^k$ represents the deleted subset from table $k$, and other tables $T^j$ (for j $\neq$ k)  remain complete.
As fully traversing these joined results to extract distributional features would be computationally expensive, we employ a \textbf{sampling-based} approach \cite{zhao2018random} that captures key distributional characteristics through limited sampling iterations $N_s$. The sampled results from each $J_d^k$ are then used to compute parameter importance scores that reflect the true impact of deletions under full-schema semantics.

The detailed procedure for Distribution Sensitivity Pruning is outlined in Algorithm \ref{alg:dsp}. The threshold $\alpha$ is distributed evenly across tables ($\alpha_k \leftarrow \alpha / K$).
Instead of one-shot pruning all deletions simultaneously, we adopt iterative pruning for each $J_d^k$ as iterative magnitude pruning (IMP) better preserves model performance than one-shot magnitude pruning (OMP) by allowing gradual adaptation and finer sensitivity estimation \cite{lee2018snip, frankle2018lottery}.
This removes parameters influenced by deletions and preserves those vital for retained data.

\begin{algorithm}[t]
	\caption{Distribution Sensitivity Pruning}
	\begin{algorithmic}[1]
		\STATE \textbf{Input:} $\theta^o$, $\{J_d^k\}_{k=1}^K$, $\{P^i\}_{i \in A}$, $\{P_r^i\}_{i \in A}$, $\alpha$, $N_s$
		\STATE \textbf{Output:} Pruned parameters $\theta^u$
		\STATE Set per-iteration pruning thresholds $\alpha_k \leftarrow \alpha / K$
		\STATE Compute $S_i = \frac{|P^i - P_r^i|}{P_r^i}$ for each $i \in A$
		\STATE Initialize sensitivity scores $\mathcal{I}_k \leftarrow 0$ for all $k$
		\FOR{each deleted join $J_d^k$ where $k = 1$ to $K$}
		\FOR{iteration $t = 1$ to $N_s$}
		\STATE Sample mini-batch $\mathbf{x}_t \sim J_d^k$
		\STATE Compute loss $\tilde{\mathcal{L}}(\mathbf{x}_t)$ using $S_i$
		\STATE Update sensitivity scores: $\mathcal{I}_k \leftarrow \mathcal{I}_k + \left(\frac{\partial \tilde{\mathcal{L}}(\mathbf{x}_t)}{\partial \theta}\right)^2$
		\ENDFOR
		\STATE Update $\theta^o$ by pruning with $\mathcal{I}_k$ and $\alpha_k$
		\ENDFOR
		\STATE \textbf{return} pruned parameters $\theta^u \leftarrow \theta^o$
	\end{algorithmic}
	\label{alg:dsp}
\end{algorithm}

\subsection{Domain Pruning}

To ensure that the model does not retain information about deleted values, we propose \textbf{Domain Pruning} that explicitly removes model support for completely eliminated values.

Formally, for a column $A^i$, let $Dom(A^i)$ denote the original domain and  $Dom(A_r^i)$ the retained domain after deletion. The deleted domain values $Dom(A_d^i)$ are:
\begin{equation}
	Dom(A_d^i) = Dom(A^i) \setminus Dom(A_r^i)
\end{equation}
The pruning strategy varies by column type:

\begin{itemize}
	\item \textbf{Categorical columns.} Each categorical attribute is represented via an embedding matrix $E^i$. We prune the embedding vectors corresponding to deleted domain values:
	      \begin{equation}
		      E^i_{\text{pruned}} = E^i[:, \; Dom(A_r^i)]
	      \end{equation}
	\item \textbf{Numerical columns.} After deletion, $Dom(A_r^i)$ may consist of disjoint subranges $\{[a_j, b_j]\}_{j=1}^{k}$ within the original range $[l, h]$. We transform these valid subranges into a compact continuous space:
	      \begin{equation}
		      x_{\text{pruned}} = \frac{x - a_j + \text{offset}_j}{\sum_{n=1}^{k}(b_n - a_n)} \cdot (h-l) + l'
	      \end{equation}
	      where $\text{offset}_j$ is the cumulative length of preceding intervals and $l'$ is the new lower bound. This preserves ordering within $Dom(A_r^i)$ while eliminating gaps from $Dom(A_d^i)$. Query ranges spanning deleted domains are adjusted by clamping to the nearest valid boundaries.
\end{itemize}

This input-level approach is model-agnostic and ensures eliminated domains cannot influence cardinality estimates.

\section{Experiments}

\begin{table*}[t]
	\centering
	\small
	\begin{tabular}{ccc|ccccc|ccccc}
		\toprule
		\multicolumn{3}{c|}{}                                & \multicolumn{5}{c|}{Neurocard}                & \multicolumn{5}{c}{FACE}                                                                                                                                                                       \\
		\multicolumn{3}{c|}{Deletion Ratio}                  & 0.1                                           & 0.3                      & 0.5           & 0.8           & 1             & 0.1           & 0.3           & 0.5             & 0.8            & 1                                                \\
		\midrule
		\multirow{8}{*}{\rotatebox[origin=c]{90}{Stale}}     & \multirow{4}{*}{\rotatebox[origin=c]{90}{OQ}} & 50th                     & 1.42          & 1.46          & 1.66          & 1.77          & 2.50          & 1.23            & 1.30           & 1.41           & 1.68           & 2.58           \\
		                                                     &                                               & 75th                     & 2.81          & 2.76          & 2.89          & 2.93          & 5.92          & 2.14            & 2.16           & 2.45           & 3.17           & 8.55           \\
		                                                     &                                               & 95th                     & 6.17          & 6.75          & 8.13          & 8.94          & 1.31e+5       & 8.26            & 8.66           & 10.18          & 11.52          & 2.60e+9        \\
		                                                     &                                               & 99th                     & 9.52          & 10.30         & 12.63         & 15.71         & 2.86e+7       & 35.37           & 33.49          & 402.32         & 63.36          & 3.18e+10       \\
		\cmidrule{2-13}
		                                                     & \multirow{4}{*}{\rotatebox[origin=c]{90}{CQ}} & 50th                     & 1.38          & 1.51          & 1.58          & 1.84          & 2.48          & 1.20            & 1.24           & 1.40           & 1.61           & 2.57           \\
		                                                     &                                               & 75th                     & 2.77          & 2.34          & 2.89          & 2.65          & 5.63          & 2.26            & 2.22           & 2.48           & 3.10           & 8.15           \\
		                                                     &                                               & 95th                     & 4.91          & 5.11          & 5.91          & 6.49          & 1.30e+5       & 8.29            & 8.70           & 9.96           & 11.35          & 2.61e+8        \\
		                                                     &                                               & 99th                     & 7.44          & 10.30         & 8.91          & 10.01         & 2.89e+7       & 33.93           & 32.52          & 437.10         & 63.13          & 3.19e+10       \\
		\midrule

		\multirow{8}{*}{\rotatebox[origin=c]{90}{Retrain}}   & \multirow{4}{*}{\rotatebox[origin=c]{90}{OQ}} & 50th                     & 1.35          & 1.33          & 1.46          & 1.49          & 1.43          & 1.18            & 1.18           & \textbf{1.15 } & \textbf{1.13 } & 1.14           \\
		                                                     &                                               & 75th                     & 2.33          & 2.28          & 2.42          & 2.33          & 2.47          & 2.08            & 2.00           & 2.08           & 2.16           & 5.86           \\
		                                                     &                                               & 95th                     & 3.92          & \textbf{4.31} & \textbf{5.31} & 5.62          & 4.86          & 7.42            & 7.84           & \textbf{8.39}  & \textbf{7.18}  & 2.56e+5        \\
		                                                     &                                               & 99th                     & 7.21          & \textbf{5.76} & 8.18          & 9.74          & 6.59          & 42.32           & 53.83          & 247.46         & 31.18          & 9.17e+7        \\
		\cmidrule{2-13}
		                                                     & \multirow{4}{*}{\rotatebox[origin=c]{90}{CQ}} & 50th                     & 1.32          & 1.48          & 1.48          & 1.59          & 1.61          & 1.20            & \textbf{1.17}  & \textbf{1.15}  & \textbf{1.13}  & 1.20           \\
		                                                     &                                               & 75th                     & 2.19          & 2.16          & 2.50          & 2.43          & 2.76          & 2.09            & 2.06           & 2.11           & 2.17           & 6.13           \\
		                                                     &                                               & 95th                     & 4.74          & \textbf{4.11} & \textbf{3.39} & \textbf{3.73} & 5.02          & 7.77            & 7.75           & 8.79           & 7.57           & 2.89e+5        \\
		                                                     &                                               & 99th                     & 6.69          & \textbf{5.45} & \textbf{4.67} & 7.68          & 6.64          & 44.16           & 54.19          & 262.39         & 31.59          & 1.38e+8        \\

		\midrule
		\multirow{8}{*}{\rotatebox[origin=c]{90}{Fine-Tune}} & \multirow{4}{*}{\rotatebox[origin=c]{90}{OQ}} & 50th                     & 1.37          & 1.34          & 1.35          & 1.47          & 1.42          & 1.18            & 1.18           & 1.17           & 1.19           & 1.21           \\
		                                                     &                                               & 75th                     & 2.48          & 2.23          & 2.48          & 2.31          & 2.75          & 2.09            & 2.03           & 2.13           & 2.15           & 6.81           \\
		                                                     &                                               & 95th                     & 4.81          & 4.71          & 5.63          & 4.83          & 42.01         & 6.95            & \textbf{7.08}  & 8.59           & 8.14           & 5.98e+4        \\
		                                                     &                                               & 99th                     & 6.89          & 6.75          & 8.58          & 8.13          & 5142          & 20.51           & 18.81          & 169.19         & 32.31          & 5.28e+7        \\
		\cmidrule{2-13}
		                                                     & \multirow{4}{*}{\rotatebox[origin=c]{90}{CQ}} & 50th                     & 1.49          & 1.49          & 1.50          & 1.52          & 1.65          & 1.18            & 1.18           & 1.19           & 1.17           & 1.18           \\
		                                                     &                                               & 75th                     & 2.19          & 2.26          & 2.18          & 2.85          & 3.15          & 2.06            & 2.08           & 2.11           & 2.13           & 5.38           \\
		                                                     &                                               & 95th                     & \textbf{3.72} & 4.23          & 3.56          & 4.62          & 269.69        & 6.96            & 6.73           & \textbf{8.18}  & 7.86           & 1.41e+5        \\
		                                                     &                                               & 99th                     & \textbf{5.54} & 7.37          & 5.06          & 6.49          & 3.93e+4       & 20.88           & 19.23          & 152.54         & 27.63          & 3.06e+8        \\

		\midrule
		\multirow{8}{*}{\rotatebox[origin=c]{90}{CEP(Ours)}} & \multirow{4}{*}{\rotatebox[origin=c]{90}{OQ}} & 50th                     & \textbf{1.30} & \textbf{1.31} & \textbf{1.33} & \textbf{1.33} & \textbf{1.21} & \textbf{1.09}   & \textbf{1.16}  & 1.16           & 1.15           & \textbf{1.11}  \\
		                                                     &                                               & 75th                     & \textbf{2.28} & \textbf{2.19} & \textbf{2.14} & \textbf{2.00} & \textbf{1.94} & \textbf{1.53}   & \textbf{1.67}  & \textbf{1.68}  & \textbf{1.83}  & \textbf{1.77}  \\
		                                                     &                                               & 95th                     & \textbf{4.50} & 4.51          & 5.79          & \textbf{3.83} & \textbf{4.52} & \textbf{6.56 }  & 7.37           & 8.44           & 7.90           & \textbf{8.75}  \\
		                                                     &                                               & 99th                     & \textbf{6.49} & 7.75          & \textbf{6.77} & \textbf{6.69} & \textbf{6.41} & \textbf{13.58 } & \textbf{14.05} & \textbf{67.42} & \textbf{26.03} & \textbf{65.8}  \\
		\cmidrule{2-13}
		                                                     & \multirow{4}{*}{\rotatebox[origin=c]{90}{CQ}} & 50th                     & \textbf{1.30} & \textbf{1.31} & \textbf{1.31} & \textbf{1.35} & \textbf{1.30} & \textbf{1.08}   & 1.18           & 1.15           & 1.16           & \textbf{1.08}  \\
		                                                     &                                               & 75th                     & \textbf{2.17} & \textbf{2.15} & \textbf{2.08} & \textbf{2.14} & \textbf{2.07} & \textbf{1.55}   & \textbf{1.63 } & \textbf{1.68}  & \textbf{1.82 } & \textbf{1.79}  \\
		                                                     &                                               & 95th                     & 4.21          & 4.28          & 4.77          & 4.14          & \textbf{6.13} & \textbf{5.40}   & \textbf{6.72}  & 8.45           & \textbf{7.56}  & \textbf{8.59}  \\
		                                                     &                                               & 99th                     & 7.58          & 6.82          & 5.67          & \textbf{5.82} & \textbf{7.39} & \textbf{13.66}  & \textbf{14.00} & \textbf{70.68} & \textbf{26.65} & \textbf{69.53} \\
		\bottomrule
	\end{tabular}
	\caption{Q-error performance under varying deletion ratios (0.1 to 1.0) on \texttt{A-1} task in JOB-light, evaluated on both Original Queries (OQ) and Complement Queries (CQ).}
	\label{table1}
\end{table*}

\begin{table*}[t]
	\centering
	\small
	\begin{tabular}{c|c|c|cccc|cccc}
		\toprule
		\multicolumn{3}{c|}{}                                 & \multicolumn{4}{c|}{OQ}                            & \multicolumn{4}{c}{CQ}                                                                                                                                     \\

		\multicolumn{3}{c|}{}                                 & 50th                                               & 75th                   & 95th          & 99th          & 50th          & 75th            & 95th          & 99th                                            \\
		\midrule
		\multirow{16}{*}{\rotatebox[origin=c]{90}{NeuroCard}} & \multirow{4}{*}{\rotatebox[origin=c]{90}{A-2-0.5}} & stale                  & 1.46          & 2.74          & 6.72          & 10.28           & 1.43          & 2.67          & 5.44          & 7.97            \\
		                                                      &                                                    & Retrain                & 1.55          & 2.24          & \textbf{4.09} & 9.47            & 1.47          & 2.46          & \textbf{3.35} & 7.20            \\
		                                                      &                                                    & Fine-Tune              & \textbf{1.34} & 2.22          & 4.46          & 8.95            & 1.34          & 2.42          & 3.59          & 6.84            \\
		                                                      &                                                    & \textbf{CEP(Ours)}     & \textbf{1.34} & \textbf{2.17} & 4.56          & \textbf{8.87}   & \textbf{1.25} & \textbf{2.27} & 3.60          & \textbf{6.74}   \\
		\cmidrule{2-11}
		                                                      & \multirow{4}{*}{\rotatebox[origin=c]{90}{A-2-1}}   & stale                  & 1.86          & 3.47          & 1.15e+4       & 4.70e+7         & 1.77          & 3.06          & 1.05e+4       & 4.75e+7         \\
		                                                      &                                                    & Retrain                & 1.76          & 2.58          & 6.17          & 8.31            & 1.67          & 2.83          & 5.92          & 9.45            \\
		                                                      &                                                    & Fine-Tune              & 1.42          & 2.93          & 10.56         & 1.15e+4         & 1.46          & 2.38          & 12.00         & 1.14e+4         \\
		                                                      &                                                    & \textbf{CEP(Ours)}     & \textbf{1.29} & \textbf{2.13} & \textbf{5.26} & \textbf{6.34}   & \textbf{1.28} & \textbf{2.09} & \textbf{4.86} & \textbf{6.81}   \\
		\cmidrule{2-11}
		                                                      & \multirow{4}{*}{\rotatebox[origin=c]{90}{A-6-0.5}} & stale                  & 2.15          & 4.23          & 10.53         & 22.96           & 2.30          & 4.20          & 10.94         & 18.65           \\
		                                                      &                                                    & Retrain                & 1.38          & 2.12          & 3.63          & 6.06            & \textbf{1.30} & 2.31          & 4.02          & 6.20            \\
		                                                      &                                                    & Fine-Tune              & 1.32          & 2.07          & \textbf{3.55} & \textbf{5.76}   & 1.33          & 2.26          & 3.75          & 6.64            \\
		                                                      &                                                    & \textbf{CEP(Ours)}     & \textbf{1.24} & \textbf{1.84} & 3.66          & 5.77            & 1.33          & \textbf{1.85} & \textbf{3.54} & \textbf{6.13}   \\
		\cmidrule{2-11}
		                                                      & \multirow{4}{*}{\rotatebox[origin=c]{90}{A-6-1}}   & stale                  & 30.77         & 5.63e+4       & 4.38e+6       & 1.83e+7         & 22.37         & 5.48e+4       & 4.20e+6       & 1.83e+7         \\
		                                                      &                                                    & Retrain                & 1.24          & 1.70          & 4.18          & 21.84           & 1.27          & 1.64          & 5.29          & 23.55           \\
		                                                      &                                                    & Fine-Tune              & 1.61          & 3.42          & 24.7          & 4168            & 1.47          & 3.20          & 24.7          & 4208            \\
		                                                      &                                                    & \textbf{CEP(Ours)}     & \textbf{1.02} & \textbf{1.63} & \textbf{3.72} & \textbf{4.84}   & \textbf{1.03} & \textbf{1.53} & \textbf{3.62} & \textbf{4.52}   \\
		\midrule
		\midrule

		\multirow{16}{*}{\rotatebox[origin=c]{90}{FACE}}      & \multirow{4}{*}{\rotatebox[origin=c]{90}{A-2-0.5}} & stale                  & 1.32          & 2.86          & 9.09          & 33.72           & 1.32          & 2.82          & 8.74          & 36.08           \\
		                                                      &                                                    & Retrain                & 1.20          & 2.10          & 7.56          & 33.91           & 1.19          & 2.12          & 7.17          & 32.00           \\
		                                                      &                                                    & Fine-Tune              & 1.17          & 1.99          & 7.04          & 20.79           & 1.16          & 1.98          & \textbf{7.04} & 21.97           \\
		                                                      &                                                    & \textbf{CEP(Ours)}     & \textbf{1.15} & \textbf{1.79} & \textbf{7.02} & \textbf{16.26}  & \textbf{1.13} & \textbf{1.78} & 7.13          & \textbf{15.92}  \\
		\cmidrule{2-11}
		                                                      & \multirow{4}{*}{\rotatebox[origin=c]{90}{A-2-1}}   & stale                  & 1.83          & 4.86          & 1.71e+8       & 5.18e+10        & 1.83          & 4.74          & 1.70e+8       & 5.33e+10        \\
		                                                      &                                                    & Retrain                & 1.18          & 3.30          & 5.63e+5       & 2.85e+8         & 1.21          & 3.19          & 4.92e+5       & 3.17e+8         \\
		                                                      &                                                    & Fine-Tune              & 1.23          & 3.15          & 2.64e+5       & 2.01e+8         & 1.23          & 3.03          & 2.44e+5       & 2.26e+8         \\
		                                                      &                                                    & \textbf{CEP(Ours)}     & \textbf{1.08} & \textbf{1.33} & \textbf{8.20} & \textbf{56.77}  & \textbf{1.09} & \textbf{1.33} & \textbf{8.37} & \textbf{59.58}  \\
		\cmidrule{2-11}
		                                                      & \multirow{4}{*}{\rotatebox[origin=c]{90}{A-6-0.5}} & stale                  & 1.66          & 3.38          & 11.00         & 209.72          & 1.68          & 3.29          & 11.04         & 209.96          \\
		                                                      &                                                    & Retrain                & 1.16          & 2.04          & \textbf{7.54} & 205.02          & 1.16          & 2.03          & \textbf{7.06} & 222.23          \\
		                                                      &                                                    & Fine-Tune              & 1.18          & 2.12          & 8.52          & 135.54          & 1.18          & 2.08          & 8.07          & 135.82          \\
		                                                      &                                                    & \textbf{CEP(Ours)}     & \textbf{1.14} & \textbf{1.66} & 8.55          & \textbf{101.03} & \textbf{1.14} & \textbf{1.62} & 8.66          & \textbf{104.63} \\
		\cmidrule{2-11}
		                                                      & \multirow{4}{*}{\rotatebox[origin=c]{90}{A-6-1}}   & stale                  & 12.42         & 2.04e+8       & 1.30e+10      & 2.84e+10        & 12.28         & 2.06e+8       & 1.35e+10      & 2.90e+10        \\
		                                                      &                                                    & Retrain                & 6.34          & 964           & 3.15e+6       & 4.11e+7         & 6.31          & 917.5         & 3.96e+6       & 5.40e+7         \\
		                                                      &                                                    & Fine-Tune              & 5.52          & 1333          & 8.01e+5       & 1.21e+7         & 5.22          & 1459          & 4.77e+5       & 7.93e+6         \\
		                                                      &                                                    & \textbf{CEP(Ours)}     & \textbf{1.02} & \textbf{1.13} & \textbf{8.54} & \textbf{24.70}  & \textbf{1.01} & \textbf{1.11} & \textbf{9.25} & \textbf{24.80}  \\
		\bottomrule
	\end{tabular}
	\caption{Q-error results on JOB-light with deletion ratios 0.5 and 1 on \texttt{A-2} and \texttt{A-6} tasks.}
	\label{table2}
\end{table*}

\subsection{Experimental Setup}

\noindent\textbf{Datasets and Workloads.}
We evaluate on two standard CE benchmarks: (1) IMDB \cite{leis2015good} dataset with JOB-light workload containing 70 queries across 6 relational tables, with full join resulting in approximately $2 \times 10^{12}$ tuples; (2) TPC-H \cite{tpch} with 4 tables at scale factor 10 (10GB data) and 100 randomly generated queries.

\noindent\textbf{Baselines \& Models.}
We conduct comparisons against several representative unlearning baselines, including:
\textbf{(a)} \textit{Stale}, which continues using the original model without any modification;
\textbf{(b)} \textit{Retrain}, which retrains the model on the retained data;
\textbf{(c)} \textit{Fine-Tune(FT)}, which adapts the original model by training it on the retained data for a few epochs.
We evaluate these methods on two state-of-the-art CE models: the autoregressive \textbf{NeuroCard} \cite{yang2020neurocard} and the normalizing flow model \textbf{FACE} \cite{wang2021face}.

\noindent\textbf{Evaluation Measures.}
We use \textit{Q-error}~($\max(\hat{c}/c,\, c/\hat{c})$) to measure estimation accuracy, where $\hat{c}$ and $c$ are estimated and true cardinalities.
Following prior work~\cite{kurmanji2024machine}, we do not evaluate membership inference attacks (MIAs) as generative models lack explicit memorization signals. We evaluate on two query types: (i) \textbf{Original Queries (OQ)} from the workload to verify successful data removal; (ii) \textbf{Complement Queries (CQ)} with inverted range predicates to test selective forgetting. CQ prevent the model from simply redistributing probability mass from deleted regions to unrelated areas.
We report Q-error at 50th, 75th, 95th and 99th percentiles.

\noindent\textbf{Unlearning Tasks.}
We evaluate two unlearning scenarios:
(\romannumeral1) \textbf{Attribute deletion (A)}: We remove tuples based on column values, e.g., \texttt{year} $\in$ [1999, 2010].\
(\romannumeral2) \textbf{Random deletion (R)}: Tuples are randomly selected for removal.
We vary affected tables from one to all, and apply deletion ratios ranging from $0.1$ to $1$. Each task is denoted as \texttt{[Type]-[Scope]-[Ratio]}, where \texttt{Type} indicates the deletion strategy (Attribute or Random), \texttt{Scope} refers to the number of affected tables, and \texttt{Ratio} denotes the proportion of tuples deleted per table. For example, \texttt{A-2-0.5} applies column-based deletion conditions to 2 tables, removing 50\% of tuples from each, while \texttt{R-3-0.3} randomly deletes 30\% of tuples from 3 tables.

\subsection{Main Results}

\noindent\textbf{Effect of Deletion Ratios.}
Table~\ref{table1} presents Q-error results for \texttt{A-1} (attribute deletion on 1 table) task across varying deletion ratios (0.1 to 1.0) on NeuroCard and FACE models. Fine-Tune performs comparably to Retrain at low deletion ratios but fails dramatically under full deletion (deletion ratio of 1), with 99th reaching 5142 on OQ and 3.93e+4 on CQ, indicating parameter instability in sparse regions. Retrain avoids this by removing embeddings of deleted values; however, it exhibits fundamental limitations in FACE models due to their continuous nature. Since normalizing flows operate over continuous spaces, they lack the granularity to explicitly exclude specific value sets. Our method, CEP, consistently achieves the lowest Q-error across all percentiles and deletion ratios, maintaining stable performance even under full deletion (50th at 1.21, 99th at 6.41). Interestingly, CEP occasionally outperforms full retraining. We hypothesize this is due to the model sparsification effect of distribution sensitivity pruning, echoing the lottery ticket hypothesis \cite{frankle2018lottery}. Furthermore, our experiments show that CEP enables effective and selective unlearning across different model architectures and deletion ratios, maintaining stable performance as deletion intensity increases while properly excluding deleted data regions.

\begin{figure}[t]
	\centering
	\includegraphics[width=1\columnwidth]{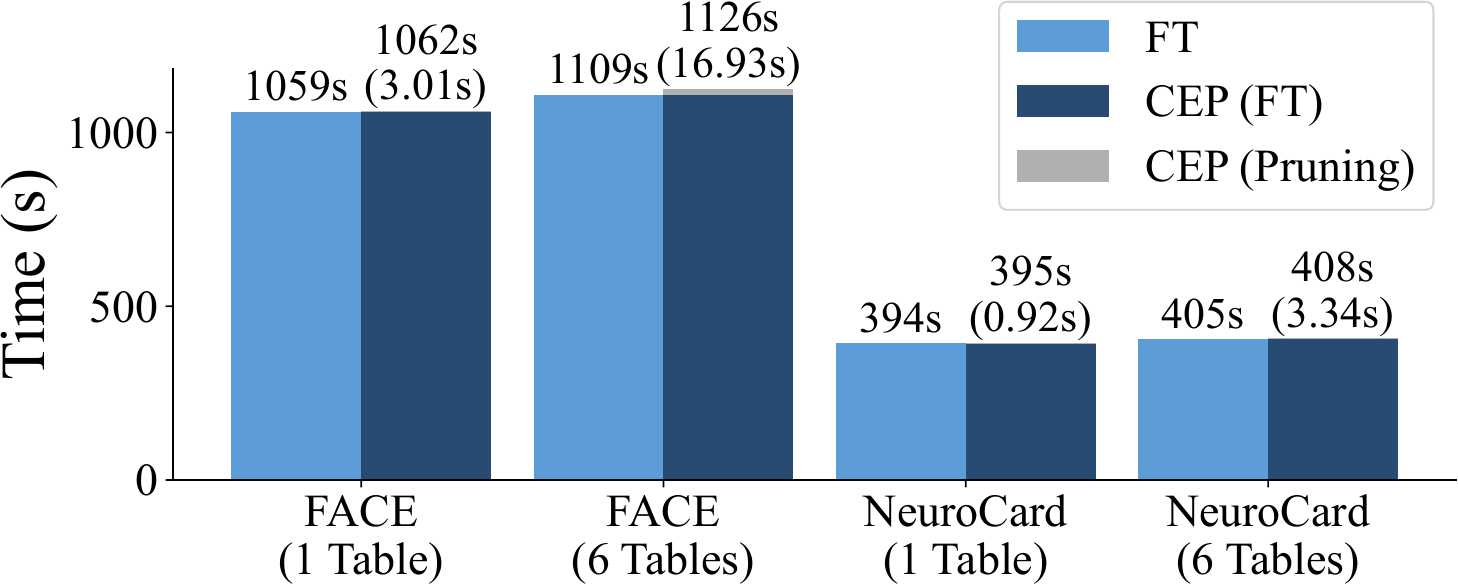}
	\caption{Training time comparison between FT and CEP on FACE and NeuroCard under 1 and 6 tables deletion. Each bar shows the total training time. For CEP, the upper bar represents the combined time of fine-tune and pruning, with the pruning time shown in parentheses.}
	\label{training-time}
\end{figure}

\begin{figure}[t]
	\centering
	\includegraphics[width=1\columnwidth]{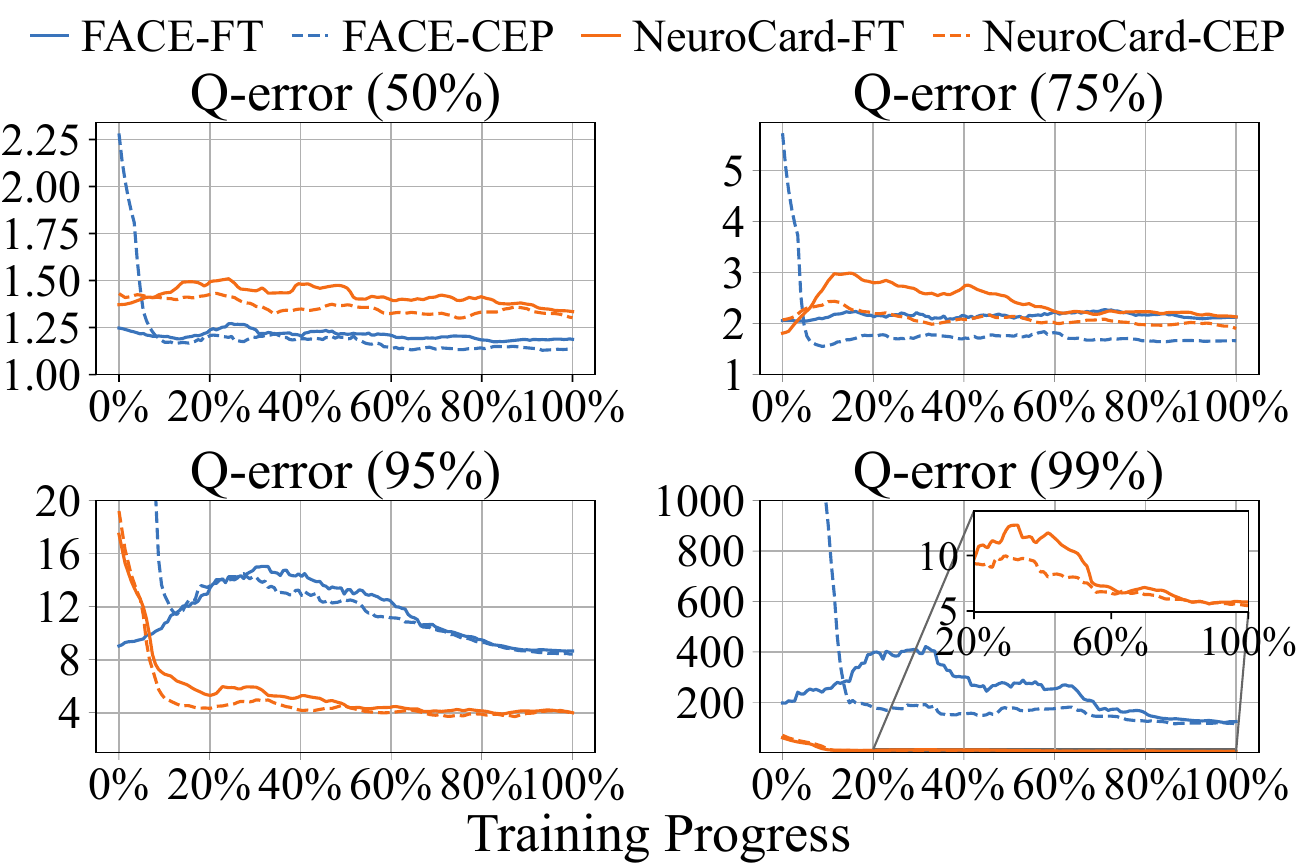}
	\caption{Training convergence curves of FT and CEP across different Q-error on \texttt{A-6-0.5} task. The x-axis is normalized to training progress (0-100\%) to enable aligned comparison between FACE and NeuroCard models.}
	\label{fig3}
\end{figure}

\noindent\textbf{Multi-Table Deletion Analysis.}
We evaluate model robustness under column deletions from \texttt{A-2} (attribute deletion on 2 tables) or \texttt{A-6} (attribute deletion on 6 tables), with deletion ratios of 0.5 and 1. On Job-light (Table~\ref{table2}), our method achieves the lowest Q-error across all deletion settings using NeuroCard and FACE. For example, in the most challenging case (A-6-1), NeuroCard achieves a 99th percentile Q-error of 4.84, significantly outperforming Retrain (21.84) and Fine-Tune (4168). On TPC-H (Table~\ref{table3}), which features more uniform data distributions, our method also maintains low Q-error across all percentiles and deletion levels, while effectively mitigating the tail error spikes observed in Fine-Tune. The consistent performance across datasets with distinct distribution characteristics, such as the skewness of Job-light and the uniformity of TPC-H, demonstrates the robustness and generalizability of our method under large-scale structural changes, while retaining better efficiency than full retraining.

\noindent\textbf{Overall Comparison.}
By comparing Table~\ref{table1} and Table~\ref{table2}, we find that CEP exhibits increasing advantages as the deletion scope expands. This demonstrates CEP's robustness across varying deletion sizes. While Fine-Tune and Retrain struggle with accuracy degradation under more extensive deletions, CEP consistently maintains low Q-error and stable performance. Notably, the relative gains of CEP are more significant in the multi-table setting, indicating its strong scalability and robustness when handling larger and more complex unlearning scenarios.

\noindent\textbf{Training Efficiency Analysis.}
To assess the efficiency and effectiveness of our proposed CEP method, we compared it against standard fine-tuning (FT) regarding total training time and convergence. Figure~\ref{training-time} shows the training time breakdown for FT and CEP under both \texttt{A-1} (attribute deletion on 1 table) and \texttt{A-6} (attribute deletion on 6 tables) tasks, using FACE and NeuroCard models. The pruning phase in CEP adds negligible overhead. For \texttt{A-1}, pruning is only 0.3\% of the FT duration. Even for \texttt{A-6}, pruning is just 2.5\% of the FT cost for the computationally intensive FACE model.
Figure~\ref{fig3} contrasts the training curves across different Q-error quantiles on \texttt{A-6-0.5}. Due to the differing training times between FACE and NeuroCard models, we normalized the x-axis to training progress (0-100\%) for a unified comparison. The results show that CEP consistently achieves faster convergence than FT, indicating that CEP facilitates early stopping.

\noindent\textbf{Random Deletion Evaluation.}
\begin{table}[t]
	\centering
	\small
	\setlength{\tabcolsep}{1mm}
	\begin{tabular}{c|c|cccc|cccc}
		\toprule
		\multicolumn{2}{c|}{}                              & \multicolumn{4}{c|}{NeuroCard} & \multicolumn{4}{c}{FACE}                                                                                                                  \\

		\multicolumn{2}{c|}{}                              & 50th                           & 75th                     & 95th          & 99th          & 50th          & 75th          & 95th          & 99th                           \\
		\midrule
		\multirow{4}{*}{\rotatebox[origin=c]{90}{R-2-0.3}} & stale                          & 1.60                     & 3.25          & 6.84          & 13.13         & 1.20          & 2.20          & 8.04          & 18.23          \\
		                                                   & Retrain                        & 1.39                     & 2.12          & \textbf{4.54} & \textbf{6.21} & 1.15          & 2.01          & 8.09          & 13.82          \\
		                                                   & Fine-Tune                      & 1.46                     & 2.30          & 4.70          & 6.78          & 1.17          & 2.14          & \textbf{6.74} & 12.93          \\
		                                                   & CEP(Ours)                      & \textbf{1.13}            & \textbf{2.05} & 5.22          & 6.66          & \textbf{1.09} & \textbf{1.84} & 7.95          & \textbf{9.94}  \\

		\midrule
		\multirow{4}{*}{\rotatebox[origin=c]{90}{R-3-0.3}} & stale                          & 1.74                     & 4.02          & 6.87          & 13.14         & 1.18          & 2.34          & 8.18          & 18.47          \\
		                                                   & Retrain                        & 1.50                     & 2.22          & \textbf{4.36} & 8.14          & \textbf{1.17} & \textbf{2.08} & 7.04          & \textbf{15.29} \\
		                                                   & Fine-Tune                      & 1.55                     & 2.22          & 4.44          & 7.49          & 1.19          & 2.10          & 6.79          & 15.43          \\
		                                                   & CEP(Ours)                      & \textbf{1.29}            & \textbf{2.11} & 5.03          & \textbf{6.56} & \textbf{1.17} & 2.13          & \textbf{6.32} & 15.38          \\
		\bottomrule
	\end{tabular}
	\caption{Q-error under 30\% random deletion on JOB-light, evaluated on \texttt{A-1} and \texttt{A-6} tasks.}
	\label{tab:random}
\end{table}
Table~\ref{tab:random} presents the Q-error under 0.3 random deletions for \texttt{R-2} (random deletion on 2 tables) and \texttt{R-3} (random deletion on 3 tables) task on JOB-light. Our method consistently delivers the lowest 50th percentile across both NeuroCard and FACE models. At higher percentiles, since random deletion lacks targeted selectivity, distinctions between methods become less apparent. Nevertheless, our approach maintains competitiveness, attaining 99th of 6.66 on NeuroCard and 9.94 on FACE in the R-2 configuration, demonstrating overall robust performance. These results indicate that our method exhibits resilience against random deletion scenarios, effectively mitigating the degradation introduced by non-selective data removal.

\begin{table}[t]
	\centering
	\small
	\setlength{\tabcolsep}{1mm}
	\begin{tabular}{c|cccc|cccc}
		\toprule
		\multirow{2}{*}{Method} & \multicolumn{4}{c|}{NeuroCard} & \multicolumn{4}{c}{FACE}                                                                                  \\
		                        & 50th                           & 75th                     & 95th          & 99th          & 50th           & 75th & 95th          & 99th   \\
		\toprule
		CEP                     & \textbf{1.21}                  & \textbf{1.94}            & \textbf{4.52} & \textbf{6.41} & 1.11           & 1.77 & \textbf{8.75} & 65.8   \\
		CEP-D                   & 1.41                           & 2.49                     & 9.92          & 2155          & 1.19           & 6.03 & 51801         & 4.10+e5 \\
		CEP-S                   & 1.25                           & 2.10                     & 5.10          & 7.01          & 1.15           & 2.21 & 12.20         & 71.1   \\
		CEP-D-S                 & 1.42                           & 2.50                     & 21.6          & 2817          & 1.20           & 6.67 & 87212         & 6.12+e5 \\
		FT                      & 1.42                           & 2.75                     & 42            & 5142          & 1.21           & 6.81 & 59757         & 5.29+e7 \\
		FT+D                    & 1.26                           & 2.15                     & 5.55          & 9.29          & 1.13           & 3.15 & 15.22         & 73.21  \\
		Retrain                 & \multicolumn{4}{c|}{/}         & 1.14                     & 5.86          & 2.56+e5        & 9.17+e7                                         \\
		Retrain+D               & \multicolumn{4}{c|}{/}         & \textbf{1.09}            & \textbf{1.43} & 10.12         & \textbf{53.37}                                 \\
		\bottomrule
	\end{tabular}
	\caption{
		Ablation results on JOB-light with ratio = 1.0 on \texttt{A-1} task.
		\textbf{D}: Domain Pruning;
		\textbf{S}: Distribution Sensitivity Pruning.
		Variants (e.g., CEP-D, FT+D) represent ablations or combinations.
	}
	\label{tab:ablation}
\end{table}

\noindent\textbf{Ablation Study.}
We conduct an ablation study to evaluate the individual contributions of Domain Pruning (D) and Distribution Sensitivity Pruning (S) within our CEP framework. Table~\ref{tab:ablation} presents results for JOB-light with various component combinations on \texttt{A-1} (attribute deletion on 1 table) task. Our complete method (CEP) consistently delivers superior performance across all quantiles. Notably, disabling domain pruning (CEP-D) causes catastrophic degradation in tail errors: NeuroCard's 99th deteriorates from 6.41 to 2155, while FACE degrades from 65.8 to 4.10e+5, demonstrating that domain pruning is essential for handling eliminated value domains. Distribution sensitivity pruning (CEP-S) provides moderate yet consistent improvements, particularly under high-error conditions. When integrated with baselines, domain pruning proves remarkably effective: FT+D reduces FACE's 99th from 5.29e+7 to 73.21, and Retrain+D achieves 99th of 53.37. These findings validate that both pruning components are indispensable for our lightweight CEP framework.

\section{Conclusion}

We present CEP, the first unlearning framework for multi-table learned CE that enables efficient data deletion without full retraining. Our method combines distribution sensitivity pruning, which leverages sensitivity scores derived from distributional shifts to guide parameter pruning, with domain pruning that eliminates support for entirely removed value domains. Experiments on NeuroCard and FACE with IMDB and TPC-H datasets show that CEP matches or exceeds retraining accuracy, especially under high deletion ratios, while significantly reducing computational cost. These results establish CEP as a practical solution for ML systems in database environments with frequent data deletions, opening promising directions for adaptive unlearning mechanisms in real-world deployments.
Future work may extend CEP to support insertions and updates, explore finer-grained unlearning at the tuple or predicate level, and integrate CEP into full query optimizers to assess system-wide impact.

\section*{Acknowledgments}
This work is supported by projects funded by the Talent Fund of Beijing Jiaotong University (2025JBMC018) and Priority Academic Program Development of Jiangsu Higher Education Institutions.

\bibliography{references}

\newpage
\appendix
\renewcommand{\thesection}{\Alph{section}}
\renewcommand{\thesubsection}{\thesection.\arabic{subsection}}
\setcounter{secnumdepth}{3}

\section*{Appendix}

This appendix provides additional details and results supporting the main paper. It is organized into five sections:

\begin{itemize}
	\item \textbf{Appendix \ref{app:time_complexity}: Distribution Sensitivity Pruning—Time Complexity Analysis.} Analysis of the time complexity of the distribution sensitivity pruning algorithm.
	\item \textbf{Appendix \ref{app:exp_details}: Experimental Details.} Detailed description of the experimental setup, including hyperparameter configurations and unlearning task settings.
	\item \textbf{Appendix \ref{app:tpch_tasks}: TPC-H Deletion Tasks.}  Table of deletion tasks for the TPC-H dataset, included in the supplement due to limited space in the main paper.
	\item \textbf{Appendix \ref{app:pruning_thresholds}: Ablation Study on Pruning Thresholds.} Investigation of the impact of different pruning thresholds on unlearning accuracy and robustness.
	\item \textbf{Appendix \ref{app:sampling_iters}: Ablation Study on Sampling Iterations.} Examination of how sampling iterations \(N_s\) affect performance under join deletions.

\end{itemize}

\section{Distribution Sensitivity Pruning—Time Complexity Analysis}
\label{app:time_complexity}

The time complexity of the Distribution Sensitivity Pruning algorithm mainly depends on the following components:

\begin{itemize}
	\item \textbf{Computing attribute sensitivity} $S_i$: This step iterates over all attributes in $A$, where $|A| = d$ is the number of attributes (columns). This computation is $O(d)$.

	\item \textbf{Iterating over semi-join deletion results}: There are $K$ semi-join deletion results. The outer loop runs $K$ times.

	\item \textbf{Importance score accumulation}: For each semi-join deletion result $J_d^k$, the algorithm runs $N_s$ sampling iterations. In each iteration:
	      \begin{itemize}
		      \item Sampling a mini-batch $\mathbf{x}_t$ is typically $O(B)$, where $B$ is the batch size.
		      \item Computing weighted loss $\tilde{\mathcal{L}}(\mathbf{x}_t)$ depends on the model complexity; denote it as $O(C)$ per batch.
		      \item Computing gradients and accumulating importance scores requires backpropagation, with cost proportional to model parameters count $P$: $O(P)$ per batch.
	      \end{itemize}

	      Therefore, importance accumulation cost per deleted join is $O(N_s \times (B + C + P))$, which simplifies to $O(N_s \times (C + P))$ as $B$ is usually small and included in $C$.

	\item \textbf{Pruning operation}: The pruning step typically requires sorting or selecting weights by importance score. Assuming $P$ parameters, pruning cost is approximately $O(P \log P)$.
\end{itemize}

\noindent \textbf{Overall time complexity:}
\[
	O\left( d + K \times \left[ N_s \times (C + P) + P \log P \right] \right)
\]

\noindent In practice, since $d$ and $K$ are usually much smaller than $N_s$, $C$, and $P$, the dominant factor is:
\[
	O\left( K \times N_s \times (C + P) \right)
\]

\section{Experimental Details}
\label{app:exp_details}

We present the experimental setup, including hyperparameter configurations and unlearning task settings. All experiments are conducted on a Linux server with an NVIDIA RTX 4090 GPU, an AMD Ryzen 9 7950X 16-core CPU, and 128 GB of RAM, using Python 3.7.

Queries with selectivity below the model's sensitivity threshold—where the model predicts zero while the true result is non-zero—are excluded from evaluation to avoid inflated q-error that does not reflect actual model performance.

\subsection{Hyperparameter Configurations}

We perform hyperparameter tuning for each model to ensure fair comparison. For distribution sensitivity pruning, we set the pruning thresholds $\alpha = 0.5$ for NeuroCard and $\alpha = 0.3$ for FACE, sampling iterations $N_s$ = 50. All models are trained using the Adam optimizer with a batch size of 128.

\paragraph{FACE.} We use a maximum learning rate of 0.0015 and train the model for 150 epochs. The model includes 5 flow steps, each consisting of an LU linear transform and a piecewise rational quadratic (PRQ) coupling layer with 8 bins. Each coupling layer is parameterized by a residual network with two blocks (hidden size 64). We apply a cube partitioning strategy with 2 divisions per dimension and sample 4 points per cube. The spline tail bound is set to 3.

\paragraph{NeuroCard.} We set the learning rate to $1 \times 10^{-3}$ and train for 100 epochs. The model consists of 4 masked residual blocks, each containing two masked linear layers with hidden dimension 128 and a dropout rate of 0.1.

\subsection{Unlearning Task Configurations}

Each task is named as \texttt{[Type]-[Scope]-[Ratio]}, where \texttt{Type} is deletion strategy (A or R), \texttt{Scope} is the number of affected tables, and \texttt{Ratio} is omitted.
Below we list the detailed deletion conditions for representative tasks on IMDB and TPC-H datasets.

\begin{table*}[t]
	\centering
	\begin{tabular}{c|c|c|cccc|cccc}
		\toprule
		\multicolumn{3}{c|}{}                                 & \multicolumn{4}{c|}{OQ}                            & \multicolumn{4}{c}{CQ}                                                                                                                                 \\

		\multicolumn{3}{c|}{}                                 & 50th                                               & 75th                   & 95th          & 99th          & 50th          & 75th          & 95th          & 99th                                          \\
		\midrule
		\multirow{16}{*}{\rotatebox[origin=c]{90}{NeuroCard}} & \multirow{4}{*}{\rotatebox[origin=c]{90}{A-2-0.5}} & stale                  & 1.38          & 1.38          & 2.24          & 2.48          & 1.38          & 1.38          & 1.39          & 1.41          \\
		                                                      &                                                    & Retrain                & \textbf{1.00} & \textbf{1.00} & \textbf{1.00} & \textbf{1.01} & \textbf{1.00} & \textbf{1.00} & \textbf{1.00} & \textbf{1.01} \\
		                                                      &                                                    & Fine-Tune              & \textbf{1.00} & \textbf{1.00} & \textbf{1.00} & 1.02          & \textbf{1.00} & \textbf{1.00} & \textbf{1.00} & \textbf{1.01} \\
		                                                      &                                                    & \textbf{CEP(Ours)}     & \textbf{1.00} & \textbf{1.00} & \textbf{1.00} & \textbf{1.01} & \textbf{1.00} & \textbf{1.00} & \textbf{1.00} & \textbf{1.01} \\
		\cmidrule{2-11}
		                                                      & \multirow{4}{*}{\rotatebox[origin=c]{90}{A-2-1}}   & stale                  & 2.06          & 2.08          & 3.46e+5       & 1.19e+7       & 2.07          & 2.07          & 2.09          & 9.68e+5       \\
		                                                      &                                                    & Retrain                & \textbf{1.00} & \textbf{1.00} & \textbf{1.00} & \textbf{1.01} & \textbf{1.00} & \textbf{1.00} & \textbf{1.01} & \textbf{1.01} \\
		                                                      &                                                    & Fine-Tune              & \textbf{1.00} & \textbf{1.00} & 1.01          & 16.45         & \textbf{1.00} & \textbf{1.00} & \textbf{1.01} & 74.70         \\
		                                                      &                                                    & \textbf{CEP(Ours)}     & \textbf{1.00} & \textbf{1.00} & \textbf{1.00} & \textbf{1.01} & \textbf{1.00} & \textbf{1.00} & \textbf{1.01} & 1.02          \\
		\cmidrule{2-11}
		                                                      & \multirow{4}{*}{\rotatebox[origin=c]{90}{A-4-0.5}} & stale                  & 1.18          & 1.20          & 1.89          & 2.25          & 1.14          & 1.21          & 1.62          & 2.26          \\
		                                                      &                                                    & Retrain                & \textbf{1.00} & \textbf{1.00} & \textbf{1.01} & \textbf{1.01} & \textbf{1.00} & \textbf{1.00} & \textbf{1.01} & \textbf{1.01} \\
		                                                      &                                                    & Fine-Tune              & \textbf{1.00} & \textbf{1.00} & \textbf{1.01} & \textbf{1.01} & \textbf{1.00} & \textbf{1.00} & \textbf{1.01} & \textbf{1.01} \\
		                                                      &                                                    & \textbf{CEP(Ours)}     & \textbf{1.00} & \textbf{1.00} & \textbf{1.01} & \textbf{1.01} & \textbf{1.00} & \textbf{1.00} & \textbf{1.01} & \textbf{1.01} \\
		\cmidrule{2-11}
		                                                      & \multirow{4}{*}{\rotatebox[origin=c]{90}{A-4-1}}   & stale                  & 1.42          & 1.48          & 6.62e+4       & 5.23e+6       & 1.31          & 1.49          & 3.08e+5       & 1.18e+6       \\
		                                                      &                                                    & Retrain                & \textbf{1.00} & \textbf{1.00} & \textbf{1.01} & \textbf{1.01} & \textbf{1.00} & \textbf{1.00} & \textbf{1.01} & \textbf{1.01} \\
		                                                      &                                                    & Fine-Tune              & \textbf{1.00} & \textbf{1.00} & \textbf{1.01} & 15.76         & \textbf{1.00} & \textbf{1.00} & \textbf{1.01} & 21.10         \\
		                                                      &                                                    & \textbf{CEP(Ours)}     & \textbf{1.00} & \textbf{1.00} & \textbf{1.01} & \textbf{1.01} & \textbf{1.00} & \textbf{1.00} & \textbf{1.01} & 1.02          \\
		\midrule
		\midrule

		\multirow{16}{*}{\rotatebox[origin=c]{90}{FACE}}      & \multirow{4}{*}{\rotatebox[origin=c]{90}{A-2-0.5}} & stale                  & 1.02          & 1.25          & 1.60          & 1.78          & 1.00          & 1.11          & 1.25          & 1.26          \\
		                                                      &                                                    & Retrain                & \textbf{1.00} & \textbf{1.01} & \textbf{1.01} & 1.03          & \textbf{1.00} & \textbf{1.01} & \textbf{1.02} & \textbf{1.02} \\
		                                                      &                                                    & Fine-Tune              & \textbf{1.00} & \textbf{1.01} & 1.02          & 1.04          & \textbf{1.00} & \textbf{1.01} & \textbf{1.02} & \textbf{1.02} \\
		                                                      &                                                    & \textbf{CEP(Ours)}     & \textbf{1.00} & \textbf{1.01} & 1.02          & \textbf{1.02} & \textbf{1.00} & \textbf{1.01} & \textbf{1.02} & \textbf{1.02} \\
		\cmidrule{2-11}
		                                                      & \multirow{4}{*}{\rotatebox[origin=c]{90}{A-2-1}}   & stale                  & 1.02          & 1.66          & 1.67e+5       & 5.78e+6       & 1.01          & 1.25          & 1.67          & 4.59e+5       \\
		                                                      &                                                    & Retrain                & \textbf{1.00} & \textbf{1.01} & \textbf{1.02} & 5489          & \textbf{1.00} & \textbf{1.01} & \textbf{1.02} & 3.05          \\
		                                                      &                                                    & Fine-Tune              & 1.01          & \textbf{1.01} & 124.65        & 7822          & \textbf{1.00} & \textbf{1.01} & \textbf{1.02} & 8.37          \\
		                                                      &                                                    & \textbf{CEP(Ours)}     & 1.01          & \textbf{1.01} & \textbf{1.02} & \textbf{1.02} & \textbf{1.00} & \textbf{1.01} & \textbf{1.02} & \textbf{1.02} \\
		\cmidrule{2-11}
		                                                      & \multirow{4}{*}{\rotatebox[origin=c]{90}{A-4-0.5}} & stale                  & 1.06          & 1.11          & 1.58          & 1.86          & 1.07          & 1.12          & 1.87          & 1.42          \\
		                                                      &                                                    & Retrain                & \textbf{1.00} & \textbf{1.01} & 1.03          & 1.12          & \textbf{1.00} & \textbf{1.01} & 1.03          & 1.12          \\
		                                                      &                                                    & Fine-Tune              & \textbf{1.00} & \textbf{1.01} & 1.03          & 1.05          & \textbf{1.00} & \textbf{1.01} & 1.03          & 1.04          \\
		                                                      &                                                    & \textbf{CEP(Ours)}     & \textbf{1.00} & \textbf{1.01} & \textbf{1.02} & \textbf{1.04} & \textbf{1.00} & \textbf{1.01} & \textbf{1.02} & \textbf{1.04} \\
		\cmidrule{2-11}
		                                                      & \multirow{4}{*}{\rotatebox[origin=c]{90}{A-4-1}}   & stale                  & 1.14          & 1.24          & 4.49e+4       & 3.54e+6       & 1.14          & 1.26          & 2.08e+5       & 7.86e+5       \\
		                                                      &                                                    & Retrain                & \textbf{1.00} & \textbf{1.01} & 147.55        & 1.66e+4       & \textbf{1.00} & 1.01          & 1640          & 7156          \\
		                                                      &                                                    & Fine-Tune              & \textbf{1.00} & \textbf{1.01} & 155.75        & 3.48e+4       & \textbf{1.00} & 1.01          & 1885          & 2.15e+4       \\
		                                                      &                                                    & \textbf{CEP(Ours)}     & \textbf{1.00} & \textbf{1.01} & \textbf{1.01} & \textbf{1.02} & \textbf{1.00} & \textbf{1.00} & \textbf{1.01} & \textbf{1.01} \\
		\bottomrule
	\end{tabular}
	\caption{Q-error results on TPC-H with deletion ratios 0.5 and 1 on \texttt{A-2} and \texttt{A-4} tasks.}
	\label{table3}
\end{table*}

\begin{itemize}
	\item \textbf{IMDB Tasks:}
	      \begin{itemize}
		      \item \texttt{A-1}
		            \begin{itemize}
			            \item \texttt{title}: \texttt{production\_year} $\in$ [1999, 2010].
		            \end{itemize}
		      \item \texttt{A-2}:
		            \begin{itemize}
			            \item \texttt{title}: \texttt{production\_year} $\in$ [1999, 2010]
			            \item \texttt{movie\_info}: \texttt{info\_type\_id} $\in$ [40, 41]
		            \end{itemize}
		      \item \texttt{A-6}:
		            \begin{itemize}
			            \item \texttt{title}: \texttt{production\_year} $\in$ [1999, 2010]
			            \item \texttt{movie\_info}: \texttt{info\_type\_id} $\in$ [40, 41]
			            \item \texttt{cast\_info}: \texttt{role\_id} $= 10$
			            \item \texttt{movie\_companies}: \texttt{company\_id} $\in$ [10000, 15000]
			            \item \texttt{movie\_keyword}: \texttt{keyword\_id} $\in$ [10000, 15000]
			            \item \texttt{movie\_info\_idx}: \texttt{info\_type\_id} $\in$ [100, 101]
		            \end{itemize}
		      \item \texttt{R-2}:
		            \begin{itemize}
			            \item \texttt{title}: \texttt{production\_year} (full value range)
			            \item \texttt{movie\_info}: \texttt{info\_type\_id} (full value range)
		            \end{itemize}
		      \item \texttt{R-3}:
		            \begin{itemize}
			            \item \texttt{title}: \texttt{production\_year} (full value range)
			            \item \texttt{movie\_info}: \texttt{info\_type\_id} (full value range)
			            \item \texttt{cast\_info}: \texttt{role\_id} (full value range)
		            \end{itemize}
	      \end{itemize}

	\item \textbf{TPC-H Tasks:}
	      \begin{itemize}
		      \item \texttt{A-2}:
		            \begin{itemize}
			            \item \texttt{customer}: \texttt{c\_mktsegment} $\in$ [0, 1]
			            \item \texttt{orders}: \texttt{o\_orderpriority} $= 3$
		            \end{itemize}
		      \item \texttt{A-4}:
		            \begin{itemize}
			            \item \texttt{customer}: \texttt{c\_mktsegment} $\in$ [0, 1]
			            \item \texttt{orders}: \texttt{o\_orderpriority} $= 3$
			            \item \texttt{lineitem}: \texttt{l\_shipmode} $= 5$
			            \item \texttt{part}: \texttt{p\_brand} $= 12$
		            \end{itemize}
	      \end{itemize}
\end{itemize}


\section{TPC-H Deletion Tasks}
\label{app:tpch_tasks}

Table~\ref{table3} lists the deletion tasks designed for the TPC-H benchmark. The deletion conditions simulate attribute-level removals across different numbers of tables, similar to the IMDB dataset. Specifically, we construct tasks with attribute deletions on 2 and 6 tables (A-2 and A-4), under deletion ratios of 0.5 and 1. These settings evaluate the model's robustness under increasingly severe schema modifications in a dataset with relatively uniform distributions.

\section{Ablation on Pruning Thresholds}
\label{app:pruning_thresholds}

\begin{figure}[ht]
	\centering
	\includegraphics[width=1\columnwidth]{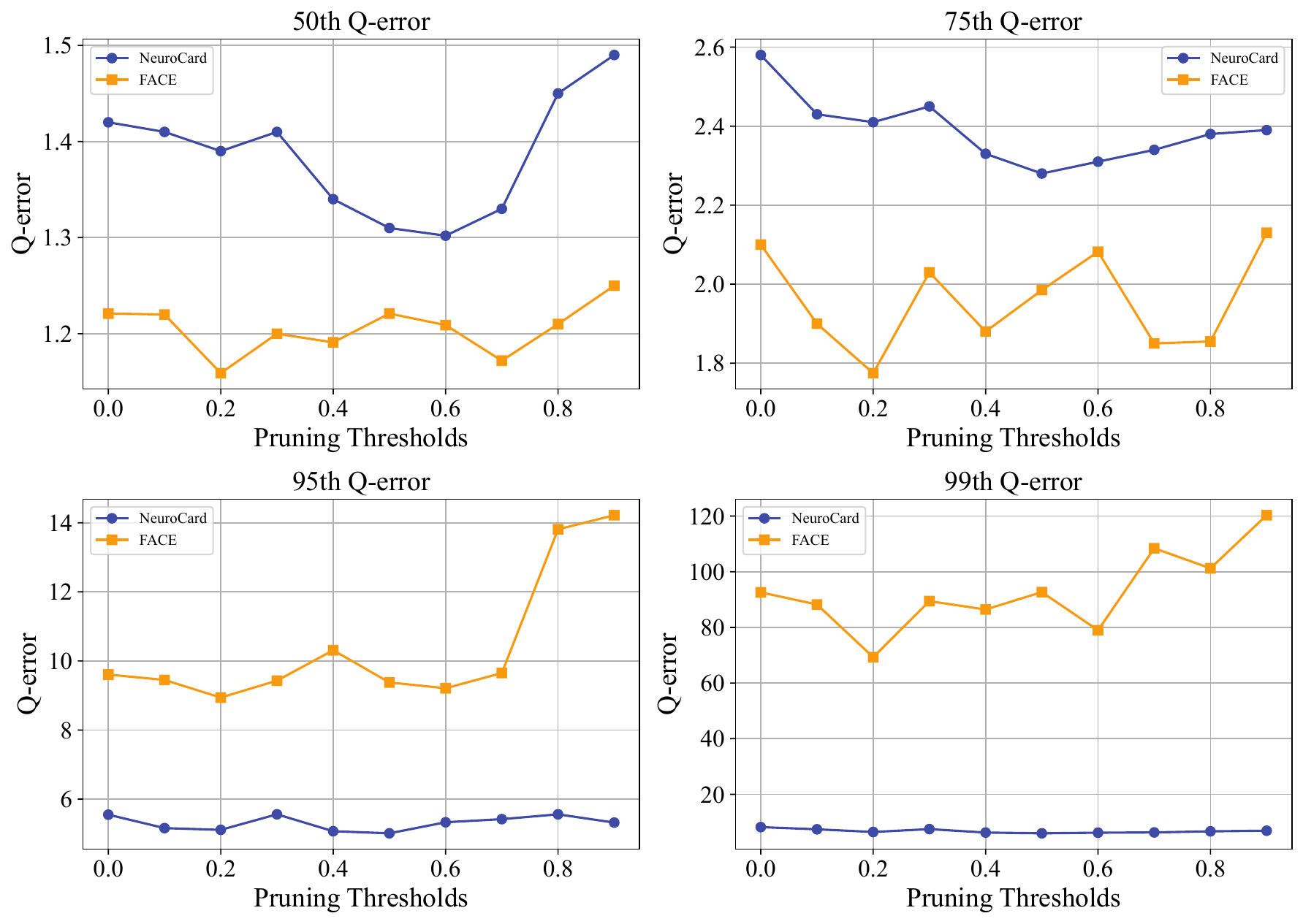}
	\caption{Ablation on pruning thresholds (A-1-0.5, JOB-light).}
	\label{fig:ablation_threshold}
\end{figure}

To study the effect of pruning strength on unlearning, we vary the pruning threshold and report Q-error across percentiles (50th-99th) on the A-1-0.5 task (JOB-light), where only distribution sensitivity pruning is applied. As shown in Figure~\ref{fig:ablation_threshold}, NeuroCard performs best at a threshold of 0.5 and remains stable under moderate pruning (0.3-0.6). In contrast, FACE achieves the lowest error at 0.2, with accuracy degrading beyond that point. These results suggest that pruning strength should be model-specific: sparse models like NeuroCard tolerate more pruning, while FACE benefits from lighter pruning to preserve high-percentile accuracy.

\section{Ablation Study on Sampling Iterations}
\label{app:sampling_iters}

\begin{figure}[ht]
	\centering
	\includegraphics[width=1\columnwidth]{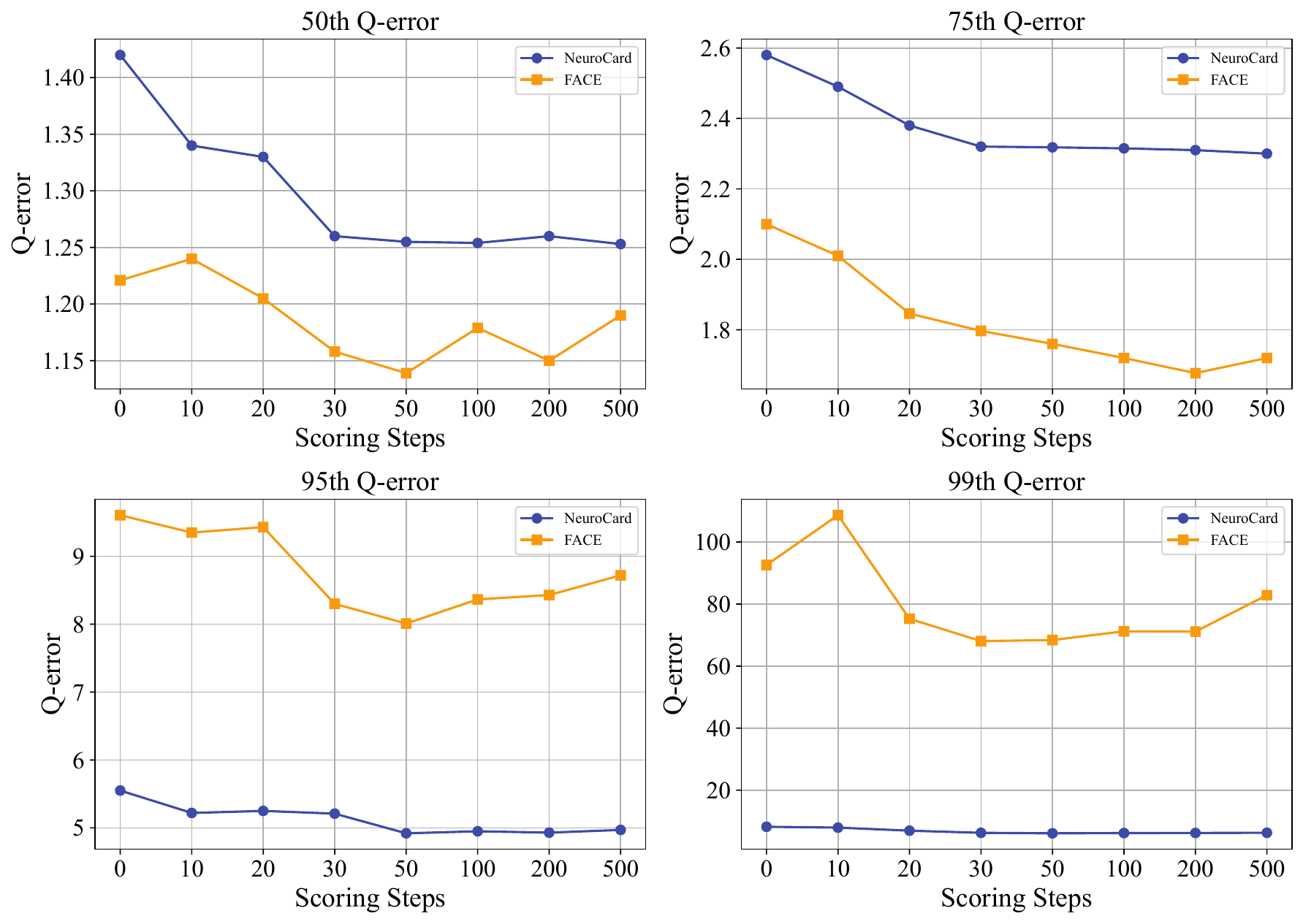}
	\caption{Ablation on sampling iterations $N_s$ (A-1-0.5, JOB-light).}
	\label{fig:ablation_steps}
\end{figure}

We study how sampling iterations $N_s$, used to accumulate parameter importance, affect pruning quality. As shown in Figure~\ref{fig:ablation_steps}, Q-error generally decreases with more steps, especially at high percentiles. NeuroCard stabilizes after 50 steps, while FACE performs best around 30 to 50 steps. These results suggest our method is sample-efficient. All main experiments use 50 sampling iterations by default.

\end{document}